\documentclass[12pt]{article}
\usepackage{amsmath,amssymb,natbib,color}
\usepackage{hyperref}
\newcommand{\unit}[1]{\,{\rm #1}}
\newcommand\arcsec{\mbox{$^{\prime\prime}$}}
\newcommand\au{\mbox{\,\textsc{au}}}
\newcommand{\cm}{\unit{cm}}
\newcommand{\gm}{\unit{g}}
\newcommand{\K}{\unit{K}}
\newcommand{\km}{\unit{km}}
\newcommand{\lsun}{\unit{L_\odot}}
\newcommand{\micron}{\unit{\mu m}}
\newcommand{\msun}{\unit{M_\odot}}
\newcommand{\pc}{\unit{pc}}
\newcommand{\rsun}{\unit{R_\odot}}
\newcommand{\s}{\unit{s}}
\newcommand{\yr}{\unit{yr}}
\newcommand{\kB}{k_{\textsc{b}}}

\newcommand{\be}{\begin{equation}}
\newcommand{\ee}{\end{equation}}

\newcommand{\sgrA}{Sgr~A$^*$}
\newcommand{\Teff}{T_{\rm eff}}
\newcommand{\Tstar}{T_{*}}
\newcommand{\HI}{{\rm HI}}
\newcommand{\HeI}{{\rm HeI}}
\newcommand{\LE}{L_{\rm E}}
\newcommand{\prad}{P_{\rm rad}}
\newcommand{\mH}{m_{\textsc{h}}}
\newcommand{\mc}{M_{\rm c}}
\newcommand{\rc}{R_{\rm c}}
\newcommand{\mJy}{\unit{mJy}}
\begin{document}
%
% TEXT GOES HERE
%
\title{On the nature of the S stars in the Galactic Center}
\author{Jeremy Goodman\thanks{{\tt jeremy@astro.princeton.edu}}
and  Bohdan Paczynski\\[1ex]
Princeton University Observatory, Princeton NJ 08544}
\date{28 March 2005}

\maketitle
\begin{abstract}
Davies and King have suggested that the bright stars observed on
short-period orbits about Sgr A$^*$ (``S stars'') are old, low-mass
stripped AGB stars rather than young, high-mass main-sequence stars.
If the observationally inferred effective temperatures and luminosities
of these stars are correct, however, then DK have
grossly overestimated the post-AGB lifetimes
and hence underestimated the production rate in steady state.
In fact, the total mass in stars stripped over the age of the Galaxy
would exceed that of the
stellar cusp bound to Sgr A$^*$.  We take issue also with Davies \&
King's estimates of the energetics involved in
capturing the stars onto their present orbits.

\end{abstract}

\section{Introduction}

The S stars with detectable orbital proper motions lie within
$\lesssim0.\arcsec5\sim 4000\au$ of \sgrA, and the brighter
ones have $K=14-15\unit{mag}$ \citep{Schoedel_etal05}.
Infrared absorption lines of \HI and \HeI in
their spectra are consistent with early main-sequence B stars
\citep{Ghez_etal03,Eisenhauer_etal05}, with implied masses
$\sim 15\msun$, luminosities $\gtrsim 10^4\lsun$,
and effective temperatures $\Teff\sim 35,000\K$.
How young massive stars might form or be captured so close
to the black hole is a fascinating question, but not one
addressed here.

\cite[henceforth DK] {Davies_King05} have suggested that the S stars
are in fact old low-mass stars that were stripped of their envelopes
after evolving onto the asymptotic giant branch (AGB).  This proposal
has several apparent advantages.  (\emph{i}) Theoretically, the
luminosities of AGB stars depend opon the masses of their cores rather
than their envelopes and exceed $10^4\lsun$ for $\mc\gtrsim 0.7\msun$
\citep{Paczynski70a}, so that, if the envelopes were sufficiently
thoroughly stripped, they could have the required $L$ \& $\Teff$.
Indeed, the central stars of planetary nebulae are just such objects.
(\emph{ii}) The challenge of explaining young, short-lived stars on
tightly-bound orbits would be circumvented.  (\emph{iii}) DK argue
that the energy required to strip the AGB stars of their envelopes
would come at the expense of the orbital energy of their cores, and
that this accounts for the strongly bound orbits.

Unfortunately, the lifetimes of DK's stars would not be $\sim
10^6\yr$, as they suggest, but more like $\sim 10^3-10^4\yr$.  The
short lifetime undercuts advantages (\emph{i}) \& (\emph{ii}), while
(\emph{iii}) has its own problems (\S\ref{discuss}).

\section{Observational constraints}

The lifetimes of the S stars are sensitive to their luminosities, so it is
important to understand the extent to which these luminosities are known.

At $K$ ($\bar \lambda=2.157\micron$),
one is observing these stars well within the Rayleigh-Jeans regime:
\[
\frac{hc}{\lambda\kB\Tstar}\approx 0.22\left(\frac{3\times
10^4\K}{\Tstar}\right).
\]
Consequently, the shape of the continuum is almost independent of the
brightness temperature $\Tstar$ and its amplitude varies only
linearly:
\begin{equation}\label{Snu}
S_\nu\approx \frac{2\pi R_*^2\kB T_* }{\lambda^2 d_{\rm GC}^2}\,.
\end{equation}
From the spectra given by \cite{Eisenhauer_etal05}, $S_\nu\sim
5-30\mJy$ at $K$; S2, which has the shortest and best-determined orbit
($15.24\pm0.36\yr$), is among the brightest: $S_\nu(S2)\approx
25\mJy$.  \cite{Eisenhauer_etal05} have corrected their spectra for an
assumed extinction of $2.8\unit{mag}$ at $K$.  The luminosity is of
course $L\equiv 4\pi\sigma\Teff^4 R_*^2$, hence very sensitive to the
actual value of $\Tstar$ if we may equate that to $\Teff$.  The
estimates for $\Teff$ rests not upon the shape of the continuum but
upon the infrared $\HI$ and $\HeI$ lines.  The implied radii and
luminosities of the S stars are
\begin{eqnarray}\label{stellar}
R_*&=& 4.6\left(\frac{S_{K}}{10\mJy}\right)^{1/2}
\left(\frac{3\times10^4\K}{\Tstar}\right)^{1/2}\rsun\nonumber\\
L_*&=& 1.5\times10^4\left(\frac{S_{K}}{10\mJy}\right)
\left(\frac{\Tstar}{3\times10^4\K}\right)^3\left(\frac{\Teff}{\Tstar}\right)^4
\lsun.
\end{eqnarray}
We have used \cite{Eisenhauer_etal05}'s value for the distance to the
Galactic Center: $d_{\rm GC}=7.62\pm0.36\unit{kpc}$, which is based
mainly on the orbit of S2.

It is clear that the main uncertainty in determining $L_*$ is the inference
of $\Teff$ from the absorption lines.  If $\Teff\sim15,000\K$ instead of
$30,000\K$, the lifetime estimated in \S\ref{theory} could be very much
longer: by almost two orders of magnitude according to the analytic
scalings, which neglect post-AGB winds.

\section{Theoretical lifetimes of shell-burning stars}\label{theory}

It is well established that the luminosity of AGB and post-AGB stars is solely
a function of the mass of the degenerate carbon-oxygen core, $\mc$,
until the envelope is exhausted.  Fundamentally, this is because the CNO and
triple-alpha reactions are very sensitive to temperature, while
hydrostatic equilibrium demands that the temperature at the base of a
radially extended envelope is proportional to $G\mc/\rc$.
\cite{Paczynski70a} quotes
\begin{equation}\label{LM_bp}
\frac{L}{\lsun}\approx 5.925\times 10^4\left(\frac{\mc}{\msun}-0.522\right)\qquad 
\mc\ge0.57\msun.
\end{equation}
The more recent calculations of \cite{Vassiliadis_Wood94} yield a similar result,
\begin{equation}\label{LM_VW}
\frac{L}{\lsun}\approx 5.6694\times 10^4\left(\frac{\mc}{\msun}-0.5\right)\qquad 
\end{equation}
On the other hand, given $L$ and hence $\mc$ and $\rc$, the
photospheric radius ($R$) and effective temperature $\Teff$ clearly
depend only on the mass and composition of the envelope.  Thus, the structure and
evolution of DK's stripped stars is the same as that of
the central stars of planetary nebulae (henceforth CSPN) at the same
$L$ and $\Teff$, notwithstanding differences in how the bulk of the
AGB envelope was lost.

Using either of relations \eqref{LM_bp} \& \eqref{LM_VW}, the core
mass corresponding to $L=10^4\lsun$ is $\mc\approx0.7\msun$.
Interpolating between \cite{Paczynski70a}'s models for $\mc=0.6\msun$
and $\mc=0.8\msun$, we find that the time required to evolve from
$\Teff=10^4\K$ to $\Teff=10^5\K$ is $\approx 1300\yr$.  The
corresponding time in \cite{Vassiliadis_Wood94}'s tracks is also
$1000-2000\yr$ (depending on metallicity) for hydrogen-burning CNSP
and $\approx 3000\yr$ for helium burning ones, but requires progenitor
masses $M_{\rm init}>1\msun$ and total ages significantly less than
the age of the Galaxy because of VW's prescriptions for mass loss in
the AGB phase.

Although computer models are needed for accurate results,
the strength of the theoretical
constraints is best appreciated from simple analytical arguments.  
There is nothing new in these except the context; see,
\emph{e.g.}, \cite{Paczynski70b}.

Energy transport through the envelopes of stripped or post-AGB stars
is radiative rather than convective (the outer convective zones of main-sequence stars,
which have higher surface gravities than those hypothesized here,
disappear at $\Teff\gtrsim 7000\K$).
From the equations of hydrostatic and radiative equilibrium,
it follows that
\begin{equation}\label{Pfrac}
\frac{d\prad}{dP}= \frac{\kappa L_r}{4\pi GM_r c}~\equiv 1-\beta\,,
\end{equation}
in which all symbols have their usual meaning.  Since the mass of the
envelope is negligible compared to that of the core, as will be seen,
and since the nuclear reactions are concentrated in narrow shells,
$M_r\approx \mc$ and $L\approx L$ at $r>\rc$.
Therefore, if the opacity
is constant (a reasonable approximation for the low values of
$\rho/T^3$ relevant here), then $\beta=1-L/\LE$ is constant, where
the Eddington luminosity
$\LE\equiv4\pi G\mc/\kappa$. Integrating \eqref{Pfrac},
one has $\prad=(1-\beta)(P-P_0)$.  In the Eddington approximation,
the constant of integration is $P_0\approx\beta a\Teff^4/6$.
At the photosphere $\tau=2/3$, $\prad=2P_0$, and
in a hydrogen shell source where $T\sim 10^{7.5}\K$,
$\prad/P_0\sim10^{12}$.
Thus $P_0$ can be
ignored for the purposes of estimating the envelope mass, so that the
gas pressure $P_{\rm gas}=\kB T\rho/\mu\mH$ and radiation pressure
$\prad=aT^4/3$ are in the constant ratio $\beta/(1-\beta)$.
The envelope is therefore polytropic:
\begin{equation}\label{polytrope}
\rho\approx \frac{\beta}{1-\beta}\frac{\mu\mH a}{3\kB}\,T^3.
\end{equation}
Using these relations to eliminate $\rho$ and $P$  in favor of $T$ in
the hydrostatic equation $dP/dr=-GM\rho/r^2$ leads to
\begin{equation}\label{structure}
T\approx \frac{G\mc\beta}{4\kB}\left(\frac{1}{r}-\frac{1}{R}\right),
\end{equation}
where $R$ is another constant of integration and is comparable to the
photospheric radius.  
Combining \eqref{polytrope} \& \eqref{structure} and integrating from
$\rc$ to $R$ yields the envelope mass:
\begin{equation}\label{menv}
M_{\rm env}\approx \frac{\pi a}{48}\left(\frac{\mu\mH}{\kB}\right)^4
(G\mc)^3\,\frac{\LE}{L_*}\left(1-\frac{L_*}{\LE}\right)^4
f\left(\frac{R}{\rc}\right),
\end{equation}
where
\begin{equation}\label{fdef}
f(x)\equiv\int\limits_1^x\left(1-\frac{t}{x}\right)^4\frac{dt}{t}~\approx~
\begin{cases}\frac{1}{4}(x-1)^4 & \ln x\ll 1,\\
\ln x & \ln x\gg 1.
\end{cases}
\end{equation}
$\rc\approx 10^{-2}\rsun$ as for a cold carbon-oxygen white dwarf, and
$R\approx 5\rsun$ to match eq.~\eqref{stellar}, so $\ln(R/\rc)\approx
6$.  For solar composition and electron-scattering opacity (larger
$\kappa$ implies smaller $M_{\rm env}$), $\kappa\approx
0.34\cm^2\gm^{-1}$ and $\mu\approx 0.62$.  Taking $\mc=0.7\msun$,
$\LE\approx 2.7\times 10^4\lsun$, $L=10^4\lsun$, \eqref{menv} predicts
$M_{\rm env}\lesssim 1.3\times 10^{-3}\msun$.  Hydrogen burning will
consume all of this in $4600\yr$, somewhat longer than the
times cited for the computer models above.  However, as the fuel is consumed,
eqs.~\eqref{menv}\&\eqref{fdef} imply that $R/\rc$ must decrease
exponentially with time.  The corresponding e-folding time of $\Teff$
is $\tau\approx1500\yr$.  If one applies the same formulae to a pure
helium envelope at the same $L$, the increased molecular weight
($4/3$) and decreased opacity ($0.2\cm^{2}\gm^{-1}$) give $M_{\rm
env}\approx 0.14\msun$, \emph{i.e.} two orders of magnitude larger.
But since the energetic yield per gram of helium fusion is about a
tenth that of hydrogen, the timescale $\tau$ increases roughly tenfold.
For smaller luminosities, $M_{\rm env}$ and $\tau$ vary
approximately as $L^{-1}$ and $L^{-2}$, respectively.

Mass loss by winds would further shorten the evolutionary timescale.

\section{Discussion}\label{discuss}

\cite{Eisenhauer_etal05} tally
$\approx 10$ stars with measured velocities $\ge 500\km\s^{-1}$ 
and $K=14-16\unit{mag}$  within $0.\arcsec7$ of \sgrA.  Adopting their
extinction $A_K=2.8\unit{mag}$, one finds that the median flux density
of these stars is $\bar S_K\approx 9\mJy$, which corresponds following
eq.~\eqref{stellar} to
$\bar L=1.3\times10^4\lsun$ if $\Teff=3\times10^4\K$.  The corresponding
median lifetime $\bar\tau\lesssim 10^4\yr$.  Thus, if the current epoch is
typical, the putative AGB stars would have to be stripped and captured
by the black hole at a rate $\gtrsim10^{-3}\yr^{-1}$; this would consume
$\gtrsim 10^7\msun$ over the age of the Galaxy.  However, the entire stellar
mass within $1.9\pc$ (approximately the cusp radius $GM_{\rm bh}/\sigma_*^2$)
is only $\approx 3\times10^6\msun$ \citep{Genzel_etal03}.

Although this is reason enough to reject DK's proposal, it has other
problems.  There is nothing special about $\Teff=3\times
10^4\K$\footnote{other that this being the minimum temperature for a
CSPN to ionize its nebula.  But that is not relevant to the S stars.}
Depending on the severity of stripping, one would expect stars of a
given core mass to be stripped down to a range of initial $\Teff$
extending to $<3\times 10^4\K$.  Since $L$ is fixed by $\mc$,
lower $\Teff$ produces brighter fluxes at $K$, so that the coolest
stripped stars would would outshine the observed S stars.  In other words,
fine tuning is required to explain the uniformity of the observed temperatures.

There are also dynamical difficulties.  DK equate the loss
of orbital energy by the core to the energy required to strip the envelope.
To leading order in the ratio $R/p$ of stellar to pericentral radius, however, 
the tidal field and the distribution of stripped material are symmetric
(quadrupolar) with respect to the core, so that the energy
required to unbind the material comes mainly at the expense of its own orbit
rather than that of the core.
One may perhaps avoid this difficulty if the stripping is preceeded by 
pericentral passages in which less violent tidal interactions 
agitate the envelope without removing it.  But then there is a further
difficulty.
Assuming that the star is much less tightly bound initially, DK's argument
implies a characteristic semimajor axis after stripping
\[
a\approx\frac{M_{\rm bh}}{M_{\rm env}} R_{\rm env},
\]
where $M_{\rm env}$ is the initial mass of the envelope \emph{before}
stripping and $R_{\rm env}$ is its virial radius, while $M\approx
3.6\times10^6\msun$ is the mass of the black hole. Taking $M_{\rm
env}\sim\msun$ and $R_{\rm env}\sim\rc$, DK are gratified to find
$a\sim 10^{-3}\pc$, in agreement with the observed semimajor axes of
the S stars.  However, these values of $M_{\rm env}$ and $R_{\rm env}$
are inconsistent.  The convective part of the envelope of the red
giant might indeed have a mass $\gtrsim0.1\msun$, but it is a
polytrope of index $3/2$ rather than $3$, so that its virial radius is
comparable to its outer radius, which is much
larger than the core radius.  The characteristic value of $a$ is
larger by the same factor.

We thank Scott Tremaine for helpful discussions, especially of the
dynamical issues.  The work was supported in part by NSF grant
AST-0307558 to JG.  This paper represents the views of its authors
only and not those of the NSF or any other part of the US Federal
Government.

\end{document}